%% file: main.tex
\documentclass[10pt, conference, compsocconf]{IEEEtran}
\usepackage{cite}
\usepackage{amsmath,amssymb,amsfonts}
\usepackage{algorithmic}
\usepackage{graphicx}
\usepackage{textcomp}
\usepackage{xcolor}
\usepackage{url}
\def\BibTeX{{\rm B\kern-.05em{\sc i\kern-.025em b}\kern-.08em
		T\kern-.1667em\lower.7ex\hbox{E}\kern-.125emX}}

\usepackage{bold-extra}
\renewcommand{\paragraph}[1]{\smallskip\noindent\textbf{\emph{#1.}}}
\usepackage[symbol]{footmisc}

\usepackage{slashbox}
\usepackage{amsthm,amsmath}
\newtheorem{theorem}{Theorem}
\usepackage{dblfloatfix}
\usepackage{paralist}
\usepackage{float}
\begin{document}
	
	\title{Irrationality, Extortion, or Trusted Third-parties:\\Why it is Impossible to Buy and Sell Physical Goods Securely on the Blockchain
	}
	
	\author{\IEEEauthorblockN{Amir Kafshdar Goharshady}
		\IEEEauthorblockA{Department of Computer Science and Engineering\\
			The Hong Kong University of Science and Technology\\
			Hong Kong, China\\
			goharshady@cse.ust.hk}}
	
	\maketitle
	
	\begin{abstract}
		Suppose that Alice plans to buy a physical good from Bob over a programmable Blockchain. Alice does not trust Bob, so she is not willing to pay before the good is delivered off-chain. Similarly, Bob does not trust Alice, so he is not willing to deliver the good before getting paid on-chain. Moreover, they are not inclined to use the services of a trusted third-party. Traditionally, such scenarios are handled by game-theoretic escrow smart contracts, such as BitHalo. In this work, we first show that the common method for this problem suffers from a major flaw which can be exploited by Bob in order to extort Alice. We also show that, unlike the case of auctions, this flaw cannot be addressed by a commitment-scheme-based approach. We then provide a much more general result: assuming that the two sides are rational actors and the smart contract language is Turing-complete, there is no escrow smart contract that can facilitate this exchange without either relying on third parties or enabling at least one side to extort the other.\footnote{This work is accepted for publication in IEEE Blockchain 2021.}
	\end{abstract}
	
	
	\input{intro}
	\input{classical}
	\input{commitment}

	\input{impossibility}
	\input{conclusion}
	
	\section*{Acknowledgments}
	
	The research was partially supported by the HKUST--Kaisa Joint Research Institute Project Grant HKJRI3A--055 and HKUST Startup Grant R9272.


\end{document}

%% file: intro.tex
\section{Introduction} \label{sec:intro}

\paragraph{Escrow} An \emph{escrow} is a contractual financial agreement between two transacting parties in which a trusted third-party, known as the escrow agent, first receives deposits from both parties and then distributes them to the parties based on predetermined rules and conditions agreed upon by the transacting parties. 

\paragraph{Sales Escrows} In this work, we are focused on the simplest form of escrow: Alice and Bob live far away and want to transact over the Internet. Alice intends to buy a physical good from Bob, which should be delivered via postal mail or some other physical channel. However, neither side trusts the other. As such, Alice will not pay unless she is assured that she will receive the good and Bob is not willing to dispatch the good unless he is assured that he will be paid. A simple solution to this situation is to use an escrow agent. Alice pays the money to the agent, who keeps it until the good is delivered. Bob delivers the good and proves this to the agent. The agent releases the funds to Bob only after verifying the delivery (either with Alice or otherwise). This is essentially how the majority of open online shopping websites function. They assume the role of an escrow agent between numerous buyers and sellers.

\paragraph{Trusted Third-Parties} The main deficiency in the method above is the presence of trusted third-parties (escrow agents). Trusted third-parties are undesirable for a variety of reasons:
\begin{compactitem}
	\item They charge exorbitant fees for their services. For example, Amazon charges between 6 to 25 percent of the good's price~\cite{amazon}.
	\item  They often do not provide anonymity or privacy and have access to the parties' personal data and transaction history. This can be used either for advertising to consumers~\cite{nytimes} or even for launching competing products~\cite{selldata}.
	\item The transactions are subject to the legal jurisdiction in which the escrow agent is located. For example, an agent located in the United States cannot facilitate an escrow if one of the transacting parties is in North Korea~\cite{nk}, even if the transaction is legal in the jurisdictions where the buyer and seller are located. Similarly, a Chinese agent cannot provide escrow services for the sale of Cannabis, which is strictly illegal in China~\cite{wen2020study}. 
\end{compactitem} 

\paragraph{Escrow on the Blockchain} The reasons above make escrows a prime target for implementation as smart contracts over programmable blockchains. Anonymity and privacy is ensured by default. The parties can transact using their pseudonym (public key) and can refrain from using the same key in more than one transaction. However, as we will see in the sequel, removing the trusted third-party entirely is not a simple task. As such, there are a variety of blockchain-based escrow services that, while considerably reducing the fees, still rely on a trusted third-party, either as the escrow agent or as a dispute resolver. There are many such protocols. Examples include LocalEthereum~\cite{lc}, descrow~\cite{descrow}, ESC9~\cite{esc9} and EscrowMyEther~\cite{eme}. Notably, on the Ethereum blockchain escrows with trusted third-parties are used more often than two-party escrows, presumably because they allow dispute resolution by a human.

\paragraph{BitHalo} With the aim of removing the trusted third-party altogether, the first strictly two-party escrow smart contract was provided by BitHalo~\cite{bithalo}. The basic mechanism is to ask both Alice and Bob to provide a deposit to the contract. Then, Bob has a predetermined amount of time to deliver the good to Alice. Finally, each party can tell the contract whether the good was delivered and the contract distributes the funds accordingly. If the two parties do not agree, the contract will burn all the funds. The idea is that the threat of losing their deposit forces/incentivizes the parties to be truthful when dealing with the contract. Although BitHalo is a Bitcoin-based protocol, variants of this idea have been implemented for programmable blockchains such as Ethereum, too. Examples include EthCrow~\cite{ethcrow} and Ethereum Escrow~\cite{ee}. The Solidity Language documentation~\cite{solidity} provides a variant, called \emph{safe remote purchase}, in which the funds are unlocked only if Alice confirms the goods are delivered, i.e.~they remain locked and are not refunded even if both parties agree that the good was not delivered. The same document mentions that ``there are multiple ways to solve this problem, but all fall short in one or the other way." It identifies the problem to be the inability of determining on-chain whether the item has truly arrived. However, as we will show in the sequel, the problem is indeed deeper and the incentives are not sufficient.

\paragraph{Extortion and Punishment} As we will see in Section~\ref{sec:classical}, a major flaw in the BitHalo-like approaches is that Bob can extort Alice by declaring to the contract that the good has been fulfilled even though it is not. At this point, Alice has two choices: either she tells the truth and both parties lose their deposit, or she plays along with the extortion, in which case Bob successfully obtains a higher payoff, but Alice herself also gets part of her deposit back. Assuming Alice is a rational agent who aims to maximize her own payoff, the right choice is to allow the extortion and not punish Bob. Surprisingly, the opposite is also possible, i.e.~Alice can extort Bob, too.

\paragraph{Data Hiding} Intuitively, the reason why Bob can extort Alice is that he can commit to his choice, i.e.~his declaration on whether the good is delivered, on the blockchain and Alice can see this choice. This changes the game for Alice and puts her in a position where the rational action is to yield to extortion. See Section~\ref{sec:classical} for more details. A similar situation exists in multi-party blind auction protocols, such as those in~\cite{solidity}, where the participants should not be able to see the choices/bids made by their competitors before making their own bid. In blind auction smart contracts, the problem is solved by relying on cryptographic primitives known as commitment schemes~\cite{brassard1988minimum,goldreich2007foundations}. In Section~\ref{sec:commitment}, we show why such a solution is not applicable to the escrow. The basic intuition is that such kind of data hiding works only if it is in the interest of the data's originator not to disclose it. However, in our case, the extortionist Bob has an active interest in disclosing and publicly committing to his choice.

\paragraph{Our Contribution} In this work, our contributions are as follows:
\begin{compactitem}
	\item We present the classical two-party escrow smart contract, together with its folklore game-theoretic analysis. This is a systematization of knowledge that is already widely available in the open-source smart contract community.
	\item We show that the folklore game-theoretic analysis is essentially wrong and makes an assumption of simultaneous action that does not hold on the blockchain.
	\item We demonstrate that the classical two-party escrow contract is vulnerable to an extortion attack.
	\item We provide a simple argument showing that the extortion problem cannot be mitigated by relying on commitment schemes and hiding the parties' choices.
	\item Finally, assuming that the parties are rational agents and the smart contract language is Turing complete, we argue that it is impossible to implement the basic sales escrow as a smart contract without trusted third-parties or vulnerability to extortion. In other words, any escrow smart contract has one of the following three demerits:
	\begin{compactitem}
		\item Assuming irrational agents who are willing to punish the other side, even if it is not in their own interest; or
		\item Relying on a third-party; or
		\item Enabling at least one of the two parties to extort the other.
	\end{compactitem}
\end{compactitem}

In summary, we illustrate that the smart contract and Dapp community is wrong in assuming that the current implementations of two-party escrows have a well-designed mechanism that incentivizes rational actors to be truthful. More shockingly, we show that the smart contracts on programmable blockchains have inherent limitations that make it impossible to implement such a contract. In a sense, this can be considered the first \emph{incontractability} result on programmable blockchains. This is particularly surprising since the Blockchain game-theoretic model of computation is quite versatile, e.g.~one can generate true (game-theoretic) randomness~\cite{DBLP:conf/icbc2/ChatterjeeGP19}, which is impossible on a Turing machine.

%% file: classical.tex
\section{Classical Escrow} \label{sec:classical}

In this section, we present the classical BitHalo-like escrow contract and the game-theoretic analysis that is often applied to it. We will then show how the analysis is incorrect and how extortion is enabled.

\subsection{Specification of a BitHalo-like Classical Escrow Contract} \label{sec:spec}

 Suppose that Alice wants to buy a physical good with price $x$ from Bob. In a classical escrow smart contract, both sides pay a deposit to the contract in the first step. Alice pays $2 \cdot x$ units and Bob pays $x$ units\footnote{In some variants, both parties pay a deposit of $2 \cdot x$. This does not make a difference in any of our analyses.}. After both deposits are received, the contract locks the funds for a predetermined amount of time. In this time, Bob should deliver the good to Alice. When this time expires, each of the two parties can call a function in the smart contract to declare their choice. Each party's choice is a single bit, denoting whether that party believes the good was delivered (1) or not delivered (0)\footnote{If a party does not declare their choice in time, the contract will assume a default choice for the party (usually 1).}. The contract then pays the parties in accordance with their choices:
\begin{compactitem}
	\item If both parties agree that the good was delivered, the contract pays $2 \cdot x$ units to Bob and $x$ units to Alice. Given that Alice's initial deposit was $2 \cdot x$ and Bob's was $x$, this is equivalent to transferring $x$ units from Alice to Bob.
	
	\item If both parties agree that the good was not delivered, then the contract refunds the original deposits, sending $x$ units to Bob and $2 \cdot x$ units to Alice.
	
	\item If the two parties do not agree, then the contract burns all the money.
\end{compactitem}

\begin{center}
\begin{tabular}{|c||c|c|c|}\hline
	\backslashbox{Alice}{Bob}
	& $0$ & $1$\\ \hline \hline
	$0$ & \backslashbox{$2 \cdot x$}{$x$}  & \backslashbox{$0$}{$0$}\\ \hline
	$1$ & \backslashbox{$0$}{$0$} & \backslashbox{$x$}{$2 \cdot x$}\\ \hline
\end{tabular}
\end{center}

\begin{figure*}
	\vspace{-1em}
	\begin{center}
	\includegraphics[keepaspectratio,scale=0.75]{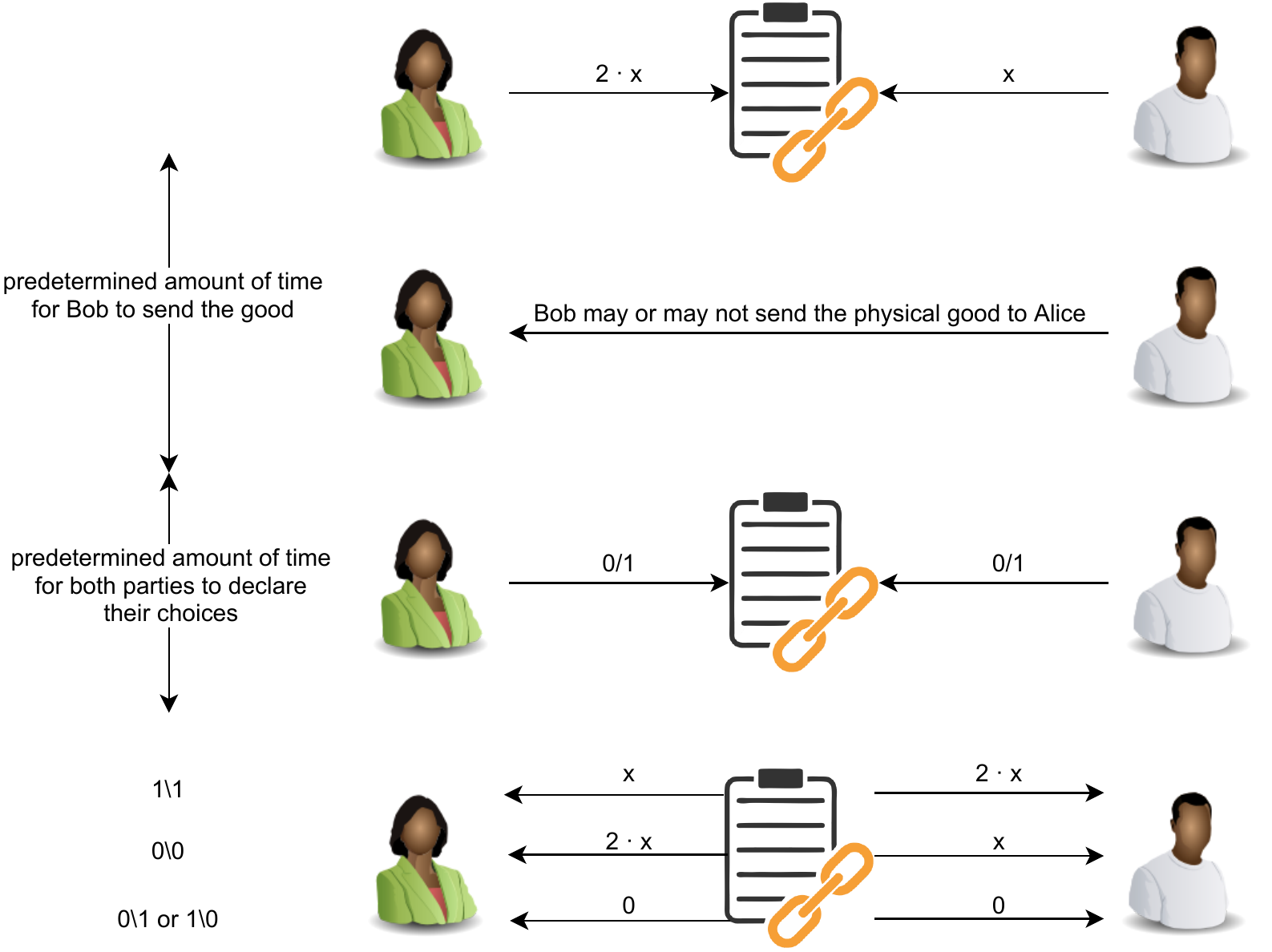}
	\end{center}
	\vspace{-1em}
	\caption{Illustration of the steps in a BitHalo-like classical escrow contract.}
	\label{fig:bithalo}
\end{figure*}

The table above shows the amount that is paid\footnote{In Ethereum, it is good practice to let the parties collect their money from the smart contract by calling a function, rather than having the smart contract proactively send the money. Throughout this paper, whenever we say a contract pays a party we assume that the payment is collected using this pattern.} to each party based on the parties' choices. An illustration of this contract is shown in Figure~\ref{fig:bithalo}. It is easy to implement this contract on any programmable blockchain and, as mentioned before, several Ethereum implementations exist.

\subsection{Folklore Game-theoretic Analysis and its Gap} \label{sec:analysis}

Although game-theoretic analysis is common in Blockchain literature~\cite{DBLP:conf/esop/ChatterjeeGV18,DBLP:conf/concur/ChatterjeeGIV18,DBLP:conf/sac/ChatterjeeGG19,zhang2019game,lewenberg2015bitcoin,johnson2014game}, we could find no formal peer-reviewed game-theoretic analysis of this escrow contract. Nevertheless, the escrow contract above, and its variants, have been used for buying and selling physical goods using various cryptocurrencies, mainly Bitcoin. The argument for its security, as presented in many informal blog posts and whitepapers, is as follows: Although the smart contract cannot access any off-chain information, both parties have access to a ground truth, i.e.~whether the physical good was actually delivered. This ground truth establishes an equilibrium in which it is in every party's best interest to declare the truth to the smart contract. In other words, the contract implements the following matrix game (normal-form game):

\begin{center}
	\begin{tabular}{|c||c|c|c|}\hline
		\backslashbox{Alice}{Bob}
		& $0$ & $1$\\ \hline \hline
		$0$ & \backslashbox[20mm]{$0$}{$0$}  & \backslashbox[20mm]{$- 2 \cdot x$}{$- x$}\\ \hline
		$1$ & \backslashbox[20mm]{$-2 \cdot x$}{$-x$} & \backslashbox[20mm]{$-x$}{$x$}\\ \hline
	\end{tabular}
\end{center}
Each player has two strategies (0 or 1) and the payoff depends on both player's choices. This is essentially the same as the matrix in Section~\ref{sec:spec}, except that the initial deposits are deducted and the numbers represent the total payoff for each player. Note that we do not consider the value of the physical good, since the payoffs in the contract are only based on the player's choices and are independent of whether the good was actually delivered.

Given the ground truth, declaring the true choice is always a correlated equilibrium:
\begin{itemize}
	\item If the ground truth is $0$, i.e.~the good is not delivered, then it is in each party's best interest to choose strategy $0$, assuming that the other party does the same. For example, knowing that Alice will choose $0$ because the good was not delivered, it is in Bob's best interest to also choose $0$ and get a payoff of $0$, rather than choosing $1$ and getting a payoff of $-x.$
	\item Similarly, if the ground truth is $1$, i.e.~the good was actually delivered in the physical world, it is in each party's interest to declare strategy $1$ to the smart contract.
\end{itemize}

As such, the folklore analysis concludes that being truthful is an equilibrium and if both parties are rational, then it is in their best interest to use the ground truth as their strategy. Note that the assumption of rationality, i.e.~that each party is motivated purely by self-interest and aims to maximize their own payoff, is central in this analysis and the contract will not work if, for example, one of the parties is malicious and aims to minimize the payoff of the other party, even if it leads to a decrease in their own payoff, too.

While the analysis of the game seems correct, a major gap in this argument is due to the fact that the game does not faithfully model the classical escrow smart contract. Specifically, in this analysis, the contract is modeled as a Nash game, i.e.~a one-shot two-player non-cooperative game with simultaneous actions taken by both players/parties. The problem is that the players' actions are essentially transactions on the blockchain, e.g.~calling functions in a smart contract. As such, they are not executed simultaneously. Indeed, as soon as one of the parties declares their choice on the blockchain, the other party can see this choice and act accordingly, i.e.~change their own choice. As we will see in Section~\ref{sec:ext}, this simple difference is precisely what enables extortion. The classical solution for mimicking simultaneous actions on the blockchain, i.e.~ensuring that each party finalizes their choice before knowing the choice of the other party, is to use a commitment scheme. We will show in Section~\ref{sec:commitment} that this solution cannot be applied for escrows.

\subsection{Extortion Attacks} \label{sec:ext}

We now demonstrate how Bob can extort Alice in the contract of Section~\ref{sec:spec}. Bob's extortion strategy is quite simple. He performs the first step of the contract and pays his deposit. However, when the funds are locked, he does not send the physical good to Alice. Then, as soon as the predetermined deadline for delivering the good expires and the parties can declare their choices, Bob sends a message (function call) to the smart contract, declaring that the good is delivered and his choice is $1$. 

At this point, Alice knows that the good is not delivered and the ground truth is $0$. However, she can also observe the blockchain and knows that Bob has declared $1$ as his choice. Hence, based on the matrix in Section~\ref{sec:analysis}, she has two choices: either she accepts Bob's lie and plays $1$, in which case Bob gets a payoff of $x$ and Alice gets $-x$, or she decides to punish Bob and play $0$, in which case Bob gets a payoff of $-x,$ but Alice herself obtains a lower payoff of $-2\cdot x.$ Clearly, if Alice is rational, she will play $1$. In other words, with the assumption of rationality, the extortion attack is successful and Bob can obtain a payoff of $x$ units without providing the physical good.

It is also noteworthy that, if Bob honestly delivers the good, then Alice has the option of extortion by declaring $0$ as her choice and leaving Bob in a similar situation, where he should either accept the extortion and get a payoff of $0$ or punish Alice but get a lower payoff of $-x.$ A rational Bob would accept the extortion, leading to Alice having the physical good without paying for it.

What the attacks above demonstrate is that the classical escrow contract does not enforce truthfulness on rational agents. Instead, it is simply a game of who gets to declare their choice first and commit it on the blockchain sooner, hence forcing the other party to concur with their choice. The order of transactions on a blockchain is essentially decided by the miners. Thus, the party with better connectivity to the miners or who is willing to pay higher transaction fees, would be able to declare their choice to the smart contract first and extort the other party.

\paragraph{Definition of Extortion} In the sequel, we assume the extortionist party is always Bob and that he is able to get his transactions on-chain before Alice. As such, we define a successful extortion as a transfer of money from Alice to Bob while the physical good is not provided.

%% file: commitment.tex
\section{Escrow with Commitment Schemes} \label{sec:commitment}

\subsection{Commitment Schemes}
When implementing blind auctions, the smart contract community has previously faced the problem of inherent lack of simultaneous actions on the blockchain. See~\cite{solidity,DBLP:conf/ithings/GoharshadyBC18,DBLP:conf/sac/ChatterjeeGP19} for some examples. In a blind auction, several parties bid for a specific item, e.g.~an NFT. However, no party should be able to know the other parties' bids before making his own bid. Otherwise, he can use this extra information to outbid the competition. 
The simple but elegant solution to this situation is to use a commitment scheme~\cite{goldreich2007foundations}. We use a smart contract with two phases. In the first phase, each party $i$ chooses their bid $b_i$ and a random nonce $n_i$. They then compute $h(b_i, n_i)$ using a predetermined one-way hash function $h$ and announce $h(b_i, n_i)$ to the contract. The contract saves the hash value in its memory. In the second phase, after all parties have already declared their hash values, each party $i$ sends the values of their bid and nonce, i.e.~$b_i$ and $n_i,$ to the contract. The contract then computes $h(b_i, n_i)$ and compares it with the hash value that is already saved. If it does not match, the bid is rejected. Otherwise, the bid is used in the auction. This is the simplest example of a commitment scheme. Basically, all parties have to commit to their bid first, without revealing it. Then, the bids will be revealed only after every party is already committed to their bid.
The fact that each party has to choose their bid and commit to it before knowing the bids of other parties effectively mimics simultaneous actions. 

\subsection{Specification of an Escrow with Commitment Schemes}
Let us now consider integrating commitment schemes into the classical escrow contract of Section~\ref{sec:spec} in order to mimic simultaneous action by Alice and Bob. The contract remains the same, except for how the choices are announced. We will have the following steps:
\begin{figure*}
	\begin{center} \vspace{-1em}
	\includegraphics[keepaspectratio, scale=0.75]{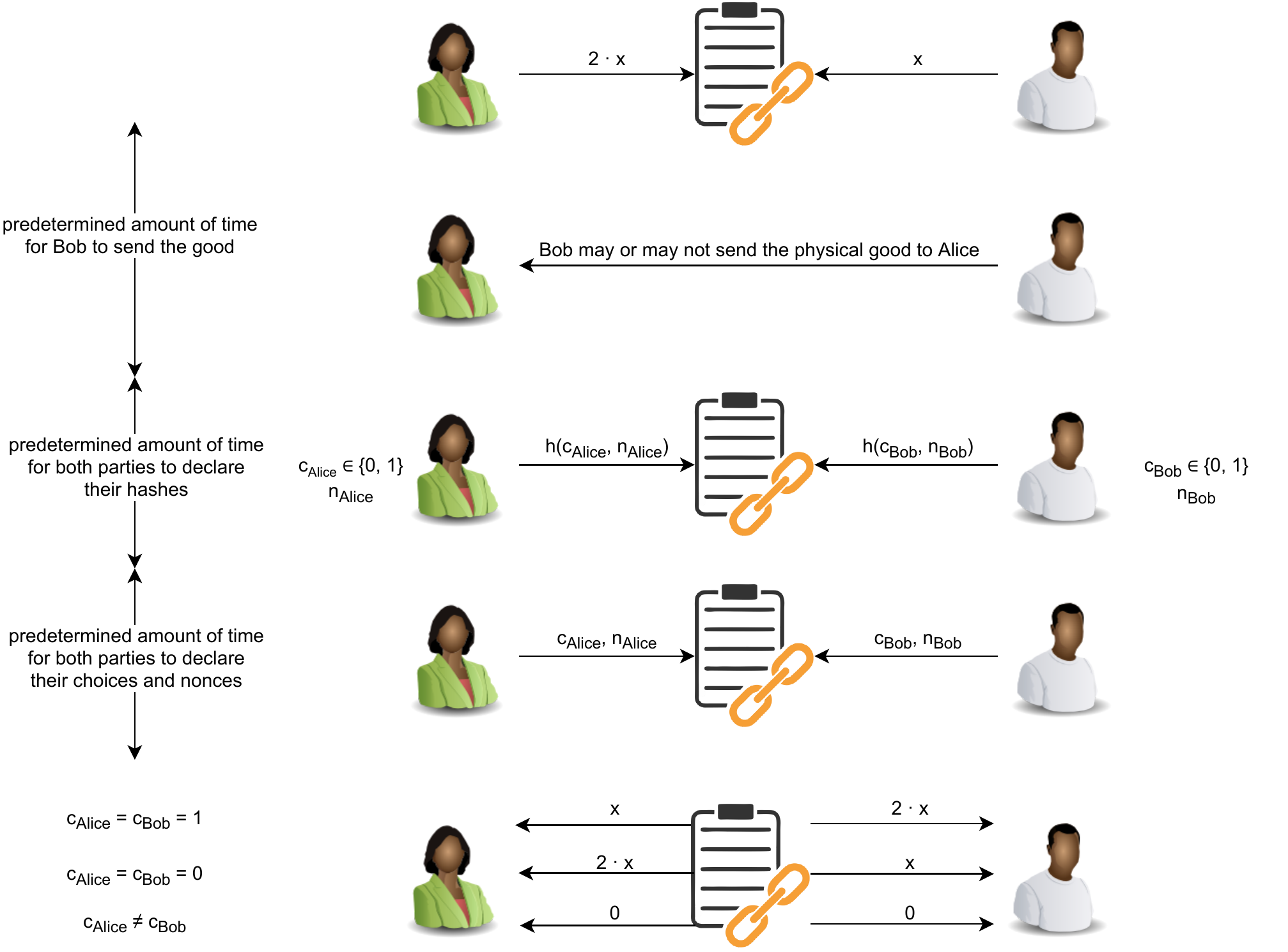}
	\end{center} \vspace{-1em}
	\caption{Illustration of the steps in an escrow contract with commitment schemes}
	\label{fig:commit}
\end{figure*}
\begin{compactenum}
	\item Alice and Bob each pay a deposit to the contract.
	\item Bob has a predetermined amount of time to physically deliver the good to Alice.
	\item At the expiry of Bob's time, both sides have a predetermined amount of time to pick their choice $c_i \in \{0, 1\}$ and a random nonce $n_i$ and announce the hash $h(c_i, n_i)$ to the contract. The contract stores this hash.
	\item After both sides have declared their hashes or the time of the previous step has passed, each party $i$ announces the values of $c_i$ and $n_i$ to the contract. The contract computes $h(c_i, n_i)$ and rejects the choice if this does not match the stored hash. Otherwise, it records the choice $c_i$ of party $i$. There is of course a time limit for this step, too. As in the classical contract, if a party fails to announce their hash, choice or nonce in time, or if there is a mismatch in the hash, a default choice will be enforced by the contract, which is usually the choice that is most favorable to the other party.
	\item The contract disburses or burns the deposit money according to the same table as in Section~\ref{sec:classical}.
\end{compactenum}
A summary of this process is illustrated in Figure~\ref{fig:commit}.

\subsection{Extortion Attack} \label{sec:comex}

At the first glance, it seems like this contract solves the problem. Since the parties are effectively acting simultaneously, the folklore game-theoretic analysis should be applicable. However, Bob can still perform an extortion attack. 

Bob's extortion strategy is as follows: He performs Step~1 and pays the deposit. In Step~2, he does not send the good to Alice. Instead, he chooses $c_{\text{Bob}} = 1$ and a nonce $n_{\text{Bob}}$ and sends a message including the pair $(c_{\text{Bob}}, n_{\text{Bob}})$ and an explanation of what they are to Alice. This message is sent via physical mail or any other delivery method that was supposed to be used for the physical good. In Step~3, he sends $h(c_{\text{Bob}}, n_{\text{Bob}})$ to the contract. Note that all transactions on the blockchain are visible to Alice. So, at this point, Alice can easily check the hash and realize that Bob has committed to a $1$ choice. As such, if Alice is a rational actor, she will be extorted just as in the previous case (Section~\ref{sec:ext}).

The subtle conceptual difference between using commitment schemes in an auction versus an escrow is that in the former the originator of a bid is incentivized to hide it in order to avoid other parties outbidding him, whereas in the latter Bob has an active interest in revealing his choice. Revealing the choice is the very action that enables him to extort Alice. As demonstrated by this example, commitment schemes can only mimic simultaneous actions when the data's originator is incentivized to hide it. This point is often overlooked by the open-source smart contract community. 

\subsection{A Defense Against the Extortion Attack}

The shrewd reader would realize that there is an easy-to-implement defense against the particular extortion attack of Section~\ref{sec:comex}. We can edit the escrow smart contract so that each party $j \in \{\text{Alice}, \text{Bob} \}$ can submit their guess of the other party's choice $c_i$ and nonce $n_i$ at any time after the hash $h_i$ is announced to the contract. The contract can of course check the correctness of the guess by simply computing $h(c_i, n_i).$ If $j$ guesses a correct $c_i$ and $n_i$ before a predetermined time and before $i$ has unmasked them, then all the funds under the contract's control will be transferred to $j$. In other words, if Bob extorts Alice by revealing his choice and nonce, then Alice can prove this extortion to the contract and obtain all the funds. 

It should be noted that this is simply a game of cat and mouse. While this defense stops the particular attack of Section~\ref{sec:comex}, there is really no need for Bob to explicitly disclose the pair $(c_{\text{Bob}}, n_{\text{Bob}})$ to Alice. All he has to do is to is to somehow prove to Alice that $c_{\text{Bob}} = 1.$ He can do this by any means. For example, he can disclose $h(c_{\text{Bob}}, n_{\text{Bob}})$ together with a witness-hiding proof~\cite{feige1990witness} showing that $c_{\text{Bob}} = 1$.
More interestingly, Bob can prove $c_{\text{Bob}} = 1$ without disclosing any information about $n_{\text{Bob}}$ by means of an auxiliary smart contract. This is the subject of Section~\ref{sec:aux} and is an idea that can be extended to any escrow contract in order to attain our incontractability result.

%% file: impossibility.tex
\section{The Auxiliary Contract Attack} \label{sec:impossibility} \label{sec:aux}

\subsection{A Particular Attack on the Contract of Section~\ref{sec:commitment}}

Consider the escrow contract with commitment schemes as presented in Section~\ref{sec:commitment}. Suppose that Bob intends to extort Alice by proving to her that his choice $c_{\text{Bob}}$ is $1$ without revealing any information about his nonce $n_{\text{Bob}}$. The basic idea in an auxiliary contract attack is that Bob creates another smart contract (an auxiliary contract) in which he loses a significant sum of money if his choice $c_{\text{Bob}}$ in the original escrow contract is not $1$. He then points Alice to this contract. Since Alice knows that Bob is a rational actor, this effectively proves to her that Bob's choice has to be $1$, which is all Bob needs to successfully extort Alice.

Let us now fill in the details of this attack. Bob's attack strategy is as follows: In Step 1, he pays the deposit to the escrow contract. In Step 2, he does not send the good to Alice. Instead, he chooses $c_{\text{Bob}} = 1$ and a nonce $n_{\text{Bob}}$ and computes the hash $h_{\text{Bob}} = h(c_{\text{Bob}}, n_{\text{Bob}}).$ He then creates an auxiliary smart contract in which $h_{\text{Bob}}$ is hard-coded. He pays this auxiliary contract a deposit of $10 \cdot x,$ which is locked by the auxiliary contract and released to Bob only if he can provide a pair $(c, n)$ such that $c = 1$ and $h(c, n) = h_{\text{Bob}}.$ In other words, Bob can get his auxiliary deposit of $10 \cdot x$ back only when he can prove to the auxiliary smart contract that $c_{\text{Bob}} = 1.$ He then sends a pointer to this auxiliary smart contract, together with an explanation of what it is, to Alice. This is delivered by the same means that was supposed to be used for the physical good. In Step~3, Bob declares $h_{\text{Bob}}$ to the escrow contract as his hash. At this point, Alice can see the auxiliary smart contract and the hash $h_{\text{Bob}}$ that Bob has declared to the escrow contract. Based on this information, she knows that Bob will lose $10 \cdot x$ units in the auxiliary contract if $c_{\text{Bob}} \neq 1.$ Given that Bob is a rational actor and would not want to lose $10 \cdot x$ units, Alice concludes that it must be the case that $c_{\text{Bob}} = 1.$ As argued in the previous cases, this is enough for a rational Alice to yield to the extortion and declare a hash corresponding to $c_{\text{Alice}} = 1.$

\subsection{A General Attack on any Escrow Contract}

We now show our most general result: On any programmable blockchain with a Turing-complete smart contract language, every strictly two-party escrow contract for the sale of a physical good between rational parties Alice (the buyer) and Bob (the seller) is subject to extortion by Bob. We establish this result by extending the auxiliary contract attack above.

We identify six basic axioms that should be satisfied by every escrow contract. Our axioms are as follows:
\begin{compactenum}
	\item \textbf{Axiom of Phases}: The escrow contract consists of three phases. In the first phase, the contract is initialized, e.g.~it collects deposits from both sides and stores their public keys. In the second phase, the contract allows time for Bob to deliver the physical good to Alice. In the third phase, the two parties can interact with the contract. This interaction leads to the contract disbursing/burning the funds in a specific manner.
	\item \textbf{Axiom of Liveness}: The escrow contract ensures that every rational party will continue their interaction with the contract and will not leave the process prematurely. One way of ensuring liveness is to enforce a default behavior whenever a party fails to take an action in its predetermined time. The default behavior can assume an action for this party that is significantly to their disadvantage or it can simply pay all the funds in the contract to the other party, hence penalizing the party that did not provide actions in time. This ensures that both parties are motivated to continue interacting with the contract until the end.
	\item \textbf{Axiom of Termination}: The contract will terminate no later than a predetermined time $t$.
	\item \textbf{Axiom of Agreement}: If both sides agree that the good was delivered, then Bob should obtain a total payoff of $x$ and Alice should have a total payoff of $-x$. Similarly, if both sides agree that the good was not delivered, then they should both have a total payoff of $0.$
	
	\item \textbf{Axiom of Incentives}: Alice's total payoff is strictly less than $-x$ if the two sides do not agree on whether the good was delivered.
	
	\item \textbf{Axiom of Boundedness}: There exists a constant $m \in (0, \infty)$ such that the total payoff of each party in the escrow contract is always in the range $[-m, m].$
\end{compactenum}
We believe the six axioms above must be satisfied by every escrow contract. The nature of a sales escrow forces the phase axiom, since Bob should at some point be given time to deliver the good off-chain. Moreover, a contract that does not satisfy liveness should be considered buggy and unacceptable, since a rational party might simply stop interacting with the contract before the process is over, possibly to the detriment of the other party. Similarly, a non-terminating contract is certainly undesirable for a one-off sale. The axiom of agreement is also part of the natural definition of an escrow. If the good is provided and both sides agree about this, then of course $x$ units of currency must move from the buyer to the seller. Similarly, if both sides agree that the good was not delivered, no money should change hands. The axiom of incentives is required because if the good is delivered (which cannot be checked on-chain by the contract) then Alice's payoff must be at most $-x$ even if there is no agreement. Of course, a similar axiom for Bob should also hold in a correctly-designed escrow smart contract, but since we are considering Bob to be the attacker and his attack is based on forcing agreement, we do not need such an axiom in our reasoning. Finally, the axiom of boundedness requires that it is not possible for a party to win/lose an arbitrarily large amount of money in the contract. 

Note that we make no assumptions on the steps and process of the first phase, or the actions that are available to the parties in the third phase.

Consider Phase 3 of the escrow contract. We do not assume anything particular about this phase. So, each party may have arbitrarily many possible action choices and, based on how the escrow contract is designed, may also be able to take a sequence of actions. Let $\Sigma_\text{Alice}$ and $\Sigma_\text{Bob}$ be the set of strategies available to Alice and Bob, respectively. A strategy is simply a recipe that, given the history of the actions taken by the two sides until this point in time, produces a suggestion for the next action to be taken by the player. In other words, a strategy for a player is simply a function mapping histories of actions to actions available to this player at the current point in the contract.

 Suppose that Alice plays $\sigma_\text{Alice} \in \Sigma_\text{Alice}$ and Bob plays $\sigma_\text{Bob} \in \Sigma_\text{Bob}.$ Then Alice's utility/payoff $u_\text{Alice} (\sigma_\text{Alice}, \sigma_\text{Bob})$ is the sum of moneys paid by the contract to Alice minus the sum of deposits paid by Alice to the contract. Bob's utility $u_\text{Bob} (\sigma_\text{Alice}, \sigma_\text{Bob})$ is defined analogously.

The axiom of agreement can be formalized as follows: there exists a strategy pair $(\sigma^1_\text{Alice}, \sigma^1_\text{Bob}) \in \Sigma_\text{Alice} \times \Sigma_\text{Bob}$ that signifies both sides are agreeing that the good was delivered and a strategy pair $(\sigma^0_\text{Alice}, \sigma^0_\text{Bob}) \in \Sigma_\text{Alice} \times \Sigma_\text{Bob}$ that signifies both sides are agreeing that the good was not delivered\footnote{The strategy pairs might not be unique.}. Moreover, we have:
\begin{compactitem}
\item $
u_\text{Alice}(\sigma^1_\text{Alice}, \sigma^1_\text{Bob}) = -x,
$
\item $
u_\text{Bob}(\sigma^1_\text{Alice}, \sigma^1_\text{Bob}) = x,
$
\item $
u_\text{Alice}(\sigma^0_\text{Alice}, \sigma^0_\text{Bob}) = u_\text{Bob}(\sigma^0_\text{Alice}, \sigma^0_\text{Bob}) = 0.
$
\end{compactitem}

Based on the axioms above, Bob can extort Alice if he can prove to Alice in Phase 2 that he will play $\sigma^1_\text{Bob}.$ If Bob can provide such a proof, then Alice either plays a strategy that agrees with Bob and leads to a payoff of $-x$ for Alice and $x$ for Bob, or she plays a strategy that disagrees with Bob and gets a lower payoff according to the axiom of incentives. A rational Alice would agree with Bob and play $\sigma^1_\text{Alice},$ leading to a successful extortion\footnote{Alice might play another strategy that agrees with $\sigma^1_\text{Bob}.$ However, the payoffs will be the same.}. 

We now show how Bob can provide this proof using two auxiliary smart contracts. As usual, Bob performs Phase 1 of the escrow contract normally, paying any potential deposits that he has to pay. In Phase 2, he does not send the physical good to Alice. Instead, he forms two auxiliary smart contracts. We call these the \emph{commitment} contract and the \emph{uniqueness} contract. 

Bob pays the commitment contract a deposit of $2 \cdot m + 1$ units. The commitment contract mimics Phase 3 of the escrow contract\footnote[1]{We can create contracts that mimic/simulate other contracts because we assume our smart contract language is Turing-complete. Hence, we can always use a universal Turing machine or its variants for our simulations. In practice, creating these simulating contracts in high-level languages such as Solidity is just a matter of changing a few lines of code.}, with the following modifications:

\begin{compactitem}
	\item The commitment contract only accepts interactions with Bob, i.e.~only Bob can call its functions.
	\item Suppose an action (function call) is taken in the escrow contract. This action will be mimicked in the commitment contract if Bob provides the original function call (which was meant for the escrow contract) to the commitment contract. Note that this should necessarily include the original timestamps and signatures. For example, if Bob wants to mimic a move taken by Alice, then the commitment contract expects a function call from Bob that provides the entirety of the function call that was made by Alice, including its timestamp and Alice's signature.
	\item Bob's deposit of $2 \cdot m + 1$ units is released to him by the commitment contract if and only if he plays $\sigma^1_\text{Bob}.$
\end{compactitem}
At this point, it is clear that Bob is committed to play $\sigma^1_\text{Bob}$ in the commitment contract, or else he loses his deposit. Moreover, the deposit amount is $2 \cdot m + 1,$ so such a loss cannot possibly be mitigated by earnings in the original contract.
 However, he needs to convince Alice that he will play $\sigma^1_\text{Bob}$ in the original escrow contract, not in the commitment contract. This is the role of the second auxiliary contract, which is called the \emph{uniqueness} contract.

The uniqueness contract takes a deposit of $4 \cdot m + 3$ from Bob and locks it until a predetermined time $t' > t.$ This contract also has a module for mimicking Phase 3 of the escrow contract. While the money is locked, i.e.~until time $t',$ the uniqueness contract allows interactions by Alice. If Alice can provide two prefixes of runs of Phase 3 of the escrow contract (with correct timestamps and signatures) in which Bob took different strategies, then the uniqueness contract would pay the entirety of the $4 \cdot m + 3$ units of deposit to Alice. Otherwise, when time $t'$ passes, the deposit will be refunded to Bob. The role of the uniqueness contract is to ensure that Bob will use the same strategy in the escrow contract and the commitment contract.

Bob sends pointers to these auxiliary contracts, together with an explanation of the attack, to Alice. This is sent via the same means that was to be used for delivering the physical good. At this point, given that Bob is rational, Alice knows that Bob will play the same exact strategy in the escrow contract and the commitment contract ($\dagger$). Otherwise, Alice would be able to steal Bob's deposit from the uniqueness contract. Given that this deposit is for $4 \cdot m + 3$ units, losing it cannot be compensated by earnings in the other two contracts. So, a rational Bob would play the same strategies in both the original and the commitment contract in order to keep his uniqueness deposit. Moreover, Alice also knows that Bob must play $\sigma^1_\text{Bob}$ in the commitment contract ($\ddagger$) or else he loses his deposit to the commitment contract. Putting ($\dagger$) and ($\ddagger$) together, it is proven to Alice that Bob will play $\sigma^1_\text{Bob}$ in the escrow contract. Just as in the previous cases, this is enough for the extortion to be successful assuming that Alice is also rational.

Based on the discussion above, we have the following theorem:
\begin{theorem} \label{thm:main}
	Every strictly two-party escrow contract for the sale of a physical good that is deployed between rational parties on a Turing-complete programmable blockchain enables the seller to extort the buyer, i.e.~obtain payment without providing the physical good.
\end{theorem}

In other words, it is impossible to buy and sell physical goods securely on the blockchain. Any escrow smart contract on a Turing-complete blockchain would either  (i)~rely on a third-party or (ii)~have irrational contracting parties, i.e.~parties who are willing to sacrifice their own payoff to punish the other party, or (iii)~enable extortion.

\paragraph{Assumptions} We conclude this section by noting that every assumption made in Theorem~\ref{thm:main} was used in our attack:
\begin{compactitem}
	\item \textbf{Rationality.} If Alice is not rational, she can stick to the truth and therefore punish Bob when he lies. However, this leads to a lower payoff for herself, too. The extortion is only successful when Alice values her own payoff more than punishing Bob. Alice's rationality is also the reason why she cannot ignore Bob's extortion messages. Ignoring the messages will lead to a lower payoff and is hence irrational. Bob's rationality is also a central assumption in the auxiliary contract attack. Alice's conviction that Bob plays $\sigma^1_\text{Bob}$ is based on the assumption that Bob is rational and would not want to lose his deposits in the two auxiliary contracts.
	
	\item \textbf{Trusted Third-party.} Adding a trusted third-party as an escrow agent would of course solve the problem of extortion. There are other ways of using a trusted third-party, too. The third-party does not necessarily have to interact with the contract or even with both parties. For example, Alice can ask a third-party to receive the physical package on her behalf and give her only one bit of information: whether the contents of the package were the desired physical good. However, using a trusted third-party is undesirable as argued in Section~\ref{sec:intro}.
	
	\item \textbf{Turing-completeness.} We rely on Turing-completeness in creating auxiliary contracts (commitment and uniqueness) that simulate parts of the original contract. 
	
	\item \textbf{Axioms.} As mentioned above, the axioms are necessary parts of any sales escrow contract. We believe a contract that does not satisfy one of our axioms cannot be considered a sales escrow. Moreover, every axiom is used either explicitly or implicitly in our extortion attack: (1)~our auxiliary contracts simulate only Phase 3 of the escrow; (2--3)~we assume every pair of strategies has a well-defined payoff. This is based on both liveness and termination. Moreover, the time $t$ is used in the construction of our uniqueness contract; (4--5)~the extortion is based on an argument of rationality that depends on payoffs satisfying the axioms of agreements and incentives; (6)~the bound $m$ is used for setting the deposit amounts in both auxiliary contracts.
\end{compactitem}

%% file: conclusion.tex
\section{Conclusion} \label{sec:conclusion}
We considered the problem of forming an escrow smart contract for the sale of a physical good when the parties are unwilling to rely on a trusted third-party. We first showed that the classical BitHalo-like escrow contracts suffer from a vulnerability to extortion attacks. We also showed that the folklore analysis of these protocols has a major gap due to absence of simultaneous actions. Moreover, this gap cannot be closed by replacing simultaneous actions with commitment schemes. Finally, we proved that assuming a Turing-complete smart contract programming language and rational actors, every strictly two-party escrow contract is vulnerable to a general extortion attack using two auxiliary contracts. This is, in a sense, the first incontractability result for programmable blockchains.